\documentstyle[preprint,aps]{revtex}
\begin{document}
\bibliographystyle{unsrt}
%\draft
\preprint{   }
%
%\begin{frontmatter}
\title{The Convergence Radius of the Chiral Expansion in the Dyson-Schwinger
Approach}
\author{T.Meissner}
\address{Institute for Nuclear Theory, University of Washington, HN-12,
\\ Seattle, WA 98195, USA}
\date{\today}
\maketitle
\begin{abstract}
We determine the convergence radius $m_{conv}$
for the expansion in the current
quark mass using the Dyson-Schwinger (DS) equation of QCD in the
rainbow approximation.
Within a Gaussian form for the gluon propagator 
$D_{\mu\nu} ({\bf p}) \sim \delta_{\mu\nu} \chi^2 e^{- {{p^2} \over \Delta}}$
we find that $m_{conv}$ increases with decreasing width $\Delta$ and
increasing strength $\chi^2$.
For those values of $\chi^2$ and $\Delta$, which provide the best known
description of low energy hadronic phenomena, $m_{conv}$ lies around
$2 \Lambda_{QCD}$, which is big enough, that the chiral expansion
in the strange sector converges.
Our analysis also explains the rather low value of 
$m_{conv} \approx 50 \dots 80 \ {\text MeV}$ in the Nambu--Jona-Lasinio
model, which as itself can be regarded as a special case of the rainbow DS 
models, where the gluon propagator is a constant in momentum space.
\end{abstract}
%\pacs{  }
%\end{frontmatter}

\newpage

The chiral perturbation theory ($\chi$PT) 
has turned out to be a rather 
powerful technique for analyzing hadronic phenomena involving Goldstone
bosons ($\pi$,$K$) as well as nucleons at low energies \cite{GL1,GL2,UGM}. 
One crucial assumption in this approach is the smallness of the
chiral symmetry breaking current mass $m_0$ compared with
$\Lambda_{QCD} \approx 200 \text{MeV}$.
This is surely fulfilled in the $u$-$d$ quark sector ($m_u = 5\text{MeV}$,
$m_d = 9 \text{MeV}$) 
but questionable in case of $s$-quarks
($m_s \approx 150 \text{MeV}$) \cite{GL3}.
It is therefore important to look at the convergence of the chiral
perturbation expansion for values of quark masses in this region.
There are mainly two features, which determine the convergence of this series:
\begin{enumerate} 
\item Non-analytic terms, such as ${m_0}^{3/2}$ or $\ln{m_0}$, which arise
from Goldstone boson ($\pi$,K) loops \cite{LP}.
\item The convergence of the power series $\sum{c_n {m_0}^n}$ itself
already on the mean field level, i.e. without any boson loops \cite{CH}.
\end{enumerate}
In this paper we will focus completely on the second point.
This problem has been studied originally in the framework 
of the chiral $\sigma$ model by Carruthers and Haymaker (CH) \cite{CH}
and recently within the Nambu--Jona-Lasinio (NJL) model \cite{NAM}
by Hatsuda \cite{HAT,HK}.
The basic idea as well as the main results are practically the same 
in both cases \footnote{One should note 
that the chiral $\sigma$ model and the NJL model
are closely connected through the gradient expansion \protect\cite{EGU}.}. 
Let us take the NJL model as an example and briefly review the main points.
The convergence radius $m_{conv}$ is determined by the existence of a solution
of the classical equation of motion (gap equation) for the 
constituent quark mass $M$, which reads in Euclidean space:
\begin{equation}
M = {m_0} + 8 {N_c} G  \int^{\Lambda_{UV}}
{ {d^4 q}\over {(2 \pi)^4}} 
{M\over{q^2 + M^2}} 
\label{eq:1}
\end {equation}
where $G$ denotes the strength of the four fermion point coupling and
$\Lambda_{UV}$ the ultraviolet cutoff.		 
Eq. (\ref{eq:1}) is solved for $M$ with $m_0$ as input, which gives a 
relation $M = M(m_0)$ (c.f. fig.\ref{fig1}).
Hereby $G$ and $\Lambda_{UV}$ are fixed and adjusted to reproduce the 
experimental value of the pion decay constant $f_\pi$ as well as a 
constituent quark mass of $M = 300 \text{MeV}$ in the chiral limit 
($m_0 =0$), which can be assumed to be a physically reasonable value.
Point ($A$) in fig.\ref{fig1} corresponds to the vacuum with 
spontaneously broken chiral symmetry ($m_0 = 0$, $M = 300 \text{MeV}$).
This is the point around which the chiral expansion in $m_0$ is
performed.
As one can see the relevant region for the convergence of the series
\begin{equation}
M (m_0) = \sum_n c_n {m_0}^n
\label{eq:1a}
\end{equation}
are the negative values of $m_0$, i.e. the branch of the curve between the
points ($A$) and ($B$).
If $m_0$ is smaller than $-|m_{conv}|$ 
(point($B$), where ${{\partial m_0}\over {\partial M}} =0$)
no solution of eq.(\ref{eq:1}) exists,
that can be reached continuously from $A$. 
This means the breakdown
of the series (\ref{eq:1a}) for all values of $m_0$ with $|m_0| > |m_{conv}|$.
From fig.\ref{fig1} we can read off $m_{conv} \approx 70 \text{MeV}$.
It has been checked that this value does not depend crucially on the 
form of the regularization scheme (e.g.O(3) 
and O(4) sharp momentum cutoffs, proper time etc. \cite{MRG}).
From this it follows that in case of the NJL model the strange quark mass 
$m_s \approx 150 \text{MeV}$ lies far beyond $m_{conv}$.
Hatsuda has demonstrated the failure of the convergence of the series
(\ref{eq:1a}) explicitly for various observables such as the quark
condensate $\langle {\bar q} q \rangle$ or the nucleon sigma term
$\Sigma_{\pi N}$ \cite{HAT,HK}.

It is the aim of this paper to study this problem within the 
rainbow Dyson-Schwinger (DS) approach to QCD \cite{HN,HIG} 
(for a comprehensive review c.f.\cite{RW}).
This model has been extensively used for describing mesonic and vacuum
properties \cite{CR,BAR,PCR,RMK} and at least tentatively also for the
calculation of nucleon observables \cite{CRP,BCP,CAH,FTF,FT}.
The general form of the rainbow DS equation for the quark propagator
$S({\bf p})$ reads:
\begin{equation}
(-i) \Sigma ({\bf p}) = {4\over 3} {g_s}^2 
\int { {d^4 k}\over {(2 \pi)^4}} 
[ (i \gamma_\mu ) ( i S ({\bf k})) (i D^{\mu\nu} ({\bf p - k}) (i \gamma_\nu) ]
\label{eq:2}
\end{equation}
where $\Sigma ({\bf p})$ denotes the quark self energy defined by
\begin{equation}
S^{-1} ({\bf p}) = {{\bf p} \!\!\! /} - m_0 - \Sigma ({\bf p})
\label{eq:3}
\end{equation}
and $D^{\mu\nu} ({\bf p - k})$ the gluon propagator.
After continuation to Euclidean space time and using the Feynman gauge
with the running coupling constant $\alpha (p^2)$ 
(where ${\bf p}$ denotes the Euclidean $4$-momentum and 
$p = \sqrt{{\bf p} \cdot {\bf p}}$)
\begin{equation}
D^{\mu\nu} ({\bf p}) = \delta^{\mu\nu} D(p^2) = \delta^{\mu\nu} 
{{4 \pi}\over{{g_s}^2}} {{\alpha(p^2)}\over {p^2}}
\label{eq:4}
\end{equation}
as well as the decomposition of the self energy $\Sigma$
\begin{equation}
\Sigma({\bf p}) = 
{{\bf p} \!\!\! /} [ A(p^2) - 1 ] + B(p^2) -m_0
\label{eq:5}
\end{equation}
one obtains the coupled set of equations:
\begin{eqnarray}
p^2 [A(p^2) -1 ] & = & 
{8\over 3} {g_s}^2 
\int { {d^4 q}\over {(2 \pi)^4}} D ((p-q)^2) 
{{A(q^2) {\bf p}\cdot {\bf q}}\over {q^2 A^2(q^2) + B^2(q^2)}} 
\label{eq:6a} 
\\ 
B(p^2) & = & m_0 +
{16\over 3} {g_s}^2 
\int { {d^4 q}\over {(2 \pi)^4}} D ((p-q)^2) 
{{B(q^2) }\over {q^2 A^2(q^2) + B^2(q^2)}} 
\label{eq:6b} 
\end{eqnarray}
As soon as a form for $D(p^2)$ is specified the eqs. (\ref{eq:6a})
and (\ref{eq:6b}) define a certain model.
It is easy to convince oneself that the gap equation of the NJL (\ref{eq:1})
appears as a special case of eqs. (\ref{eq:6a}) and (\ref{eq:6b}) if the 
gluon propagator is a constant in momentum space
\begin{equation} 
D_{NJL} (p^2) = {{3 N_c}\over 2} G
\label{eq:7a}
\end{equation}
or equivalently
\begin{equation}
\alpha_{NJL} = {{3 N_c}\over 2} G {{g_s}\over{4 \pi}} p^2
\label{eq:7b}
\end{equation}
which gives a contact interaction in coordinate space with the momentum 
independent solutions $A(p^2) =1$ and $B(p^2) =M$.
From a principle point of view none of these models, i.e. none of the
gluon propagators is preferred over the other, because 
none of them can be derived from QCD. 
On the other hand the general features of QCD are more likely better 
described with a running coupling $\alpha (p^2)$, which on the
one side grows in the 
infrared region and therefore gives rise to ``confined'' quarks and on the 
other side shows asymptotic freedom in the ultraviolet region
\begin{equation}
\alpha(p^2) \stackrel{p^2 \rightarrow \,  \infty}{\sim} 
{{d\pi}\over{\ln\left( {p^2/{\Lambda_{QCD}}^2} \right )} }   
\label{eq:66}
\end{equation}
(with $d = {12\over{33 - 2 {N_f}}}$),
rather than with a form like $\alpha_{NJL} (p^2) \sim p^2$, which gets small in
the IR region but grows for high $p^2$ and has to be cut off 
at $\Lambda_{QCD}$.
In refs.\cite{CR,PCR} a superposition of a Gaussian form with strength 
$\chi^2$ and width $\Delta$ (dominating at small $p^2$) and the standard
asymptotic form (dominating at high $p^2$)
\begin{equation}
\alpha (p^2) = {{3\pi \chi^2}\over 4} \left ( {{p^2}\over{\Delta^2}}
\right )
e^{ - p^2 /\Delta} \, + \, 
{{d\pi}\over{\ln\left( \tau + {p^2/{\Lambda}^2} \right )} }   
\label{eq:8}
\end{equation}
with $\Lambda = \Lambda_{QCD} = 190 \text{MeV}$, $\chi = 1.14 \text{GeV}$,
$\Delta = 0.002 {\text{GeV}}^2$ and $\tau = 3.0$ has been shown to provide
a reasonable description of low energy hadronic phenomena.
Furthermore it turned out that the strength of $\alpha (p^2)$ in the 
IR domain is large enough that the model shows spontaneously broken
chiral symmetry, i.e. the DS equation 
(\ref{eq:6a},\ref{eq:6b}) 
has a
nontrivial solution $B(p^2) \neq 0$ if $m_0 =0$, as well as ``confinement'',
in the sense that the quark propagator $S({\bf p})$ does not have a pole at
timelike $p^2$.

For our study of the convergence of the chiral expansion in
this model we will first consider a pure Gaussian ansatz
\begin{equation}
\alpha (p^2) = {{3\pi \chi^2}\over 4} \left ( {{p^2}\over{\Delta^2}} \right )
e^{ - p^2 /\Delta} 
\label{eq:9}
\end{equation}   
and vary the strength $\chi^2$ as well as the width $\Delta$.
For the determination of $m_{conv}$ we have again to look at negative
$m_0$.
In order to do so we write the set of the DS equations 
(\ref{eq:6a},\ref{eq:6b}) in the form:
\begin{eqnarray}
p^2 [A(p) -1 ] & = & 
{4\over 3} 
\int^\infty_0 d q q^3 
{{A(q^2) \, ({\bf p q}) }\over {q^2 A^2(q) + B^2(q)}} \cdot {\cal K}_A (p,q) 
\label{eq:10a} 
\\ 
B(p)  & = & B(0) +  
{4\over 3} 
\int^\infty_0 d q q^3 
{{B(q) }\over {q^2 A^2(q) + B^2(q)}} \cdot 
[ {\cal K}_B (p,q) - {\cal K}_B (0,0) ]
\label{eq:10b}
\end{eqnarray}
where the integral kernels ${\cal K}_A$ and ${\cal K}_B$ are given by:
\begin{equation}
{\cal K}_A (p,q) = \int d\Omega 
{{\alpha (p-q)^2)}\over {(p-q)^2}} \, ({\bf p q} ) = {{3\chi^2}\over{4\Delta}}
e^{- {{p^2 + q^2}\over \Delta} } 
\left [ { 
{I_0 \left ( {{2pq}\over\Delta} \right ) + 
 I_2 \left ( {{2pq}\over\Delta} \right ) } \over 2 } - 
{ {I_1 \left ( {{2pq}\over\Delta} \right )} \over 
{ \left ( {{2pq}\over\Delta} 
\right ) } }
\right ]
\label{eq:11a}
\end{equation}
and
\begin{equation}
{\cal K}_B (p,q) = \int d\Omega 
{{\alpha (p-q)^2)}\over {(p-q)^2}}  = {3\chi^2}
e^{- {{p^2 + q^2}\over \Delta} }
{ {I_1 \left ( {{2pq}\over\Delta} \right )} \over 
{ \left ( {{2pq}\over\Delta} 
\right ) } }
\label{eq:11b}
\end{equation}
respectively. 
Hereby $I_n$ denotes the Bessel function of order $n$.
The set of integral equations (\ref{eq:10a}) and (\ref{eq:10b}) is 
formally independent of $m_0$ but instead depends on the initial value
$B(0)$.
It can be uniquely solved for any given $B(0)$ and the corresponding $m_0$
can then be extracted from the solution $B(p)$ at high $p$ 
\begin{equation}
\lim_{p \to \infty} B(p) = m_0
\label{eq:12}
\end{equation}
This renders a relationship between $B(0)$ and $m_0$, 
i.e. $m_0 = m_0 [B(0)]$, analogous to the 
one displayed in fig.\ref{fig1}. 
The convergence radius $m_{conv}$ is then determined by the
minimum of this curve, i.e. the point:
\begin{equation}
{{\partial m_0}\over {\partial B(0) }}  = 0
\label{eq:min}
\end{equation}

The set of eqs.(\ref{eq:10a}) and (\ref{eq:10b}) is solved by a selfconsistent
procedure in the interval $[0,1000 \Lambda ]$ with logarithmic grid
points.

\begin{enumerate}
\item Let us first look at solutions with various widths within the
Gaussian parameterization (\ref{eq:9}) keeping $\chi^2$ fixed at
$\chi = 8.0 \Lambda$.The forms of the corresponding running coupling
constants $\alpha (p)$ are displayed in fig.\ref{fig2}, where for comparison
we have included the one of the NJL eq.(\ref{eq:7b}) in addition.
In fig.\ref{fig3} we compare the solutions $B(p)$ for a fixed value of
$B(0) = 7.75$, whereas fig.\ref{fig4} shows the dependence 
$m_0 = m_0 [B(0)]$.
It is interesting to look at the limes $\Delta \to 0$, which means, in fact,
that the gluon propagator approaches a $\delta$-function in momentum space
(dotted line in fig.\ref{fig4}):
\begin{equation}
D( {\bf p- q} ) 
= (2 \pi)^4 {3\over{16}} {1\over{g_s}^2} \chi^2 \delta^{(4)} ({\bf p -q} )
\label{eq:12a}
\end{equation}
In this case we obtain an algebraic relation between $m_0$ and $B(0)$ from
eq.(\ref{eq:6b}):
\begin{equation}
m_0 = B(0) - {{\chi2}\over{B(0)}}
\label{eq:13}
\end{equation}
From figs.\ref{fig3} and \ref{fig4} we can clearly see, that $m_{conv}$
rises with decreasing $\Delta$.
For small values of $\Delta$ 
(e.g. $\Delta = 0.2 \, \Lambda^2$, full line in fig.\ref{fig4})
one faces the problem that the self
consistent iteration procedure for the solution of the system 
(\ref{eq:10a}) and (\ref{eq:10b}) does not converge but ends up switching
between two or more configurations, if the value of $B(0)$ deviates
too much from the one in the chiral limit
($m_0 =0$). This is a typical feature encountered in many nonlinear systems
and does not mean that there exists no solution at all, but only that the
simple selfconsistent procedure is not able to find it.
It might be possible that more elaborate methods are successful in this case.
On the other side we are clearly able to give at least a minimum value 
for $m_{conv}$ from the corresponding curve in fig.\ref{fig4}, which is
sufficient for our analysis.
Furthermore we recognize that for those small $\Delta$ the curve, 
as far as it can be calculated from the self consistent procedure, lies very
close to the one obtained with a $\delta$-function gluon propagator
($\Delta =0$, dotted line in fig.\ref{fig4}). 
\item
For $\Delta =0$ it is clear from eq.(\ref{eq:13}) that 
\begin{equation}
\left \lbrace 
m_0 [B(0)] \right \rbrace_{\chi_1} > 
\left \lbrace m_0 [B(0)] \right \rbrace_{\chi_2}
\label{eq:14}
\end{equation}
if $\chi_1 < \chi_2$.
We have checked that this relation holds generally if the value of $\Delta$.
From this we conclude that $m_{conv}$ increases with increasing strength 
$\chi^2$.
\item Finally we have considered also gluon propagators of the form
(\ref{eq:8}), which have in addition to the Gaussian form in the IR region
have the logarithmic tail at high $p^2$ leading to asymptotic freedom.
In this case the momentum integrals in 
eqs.(\ref{eq:6a}),(\ref{eq:6b}), (\ref{eq:10a}) and (\ref{eq:10b}) 
run from $0$ to the ``renormalization
point'' $\mu^2$ and the $m_0$ is considered as the running mass $m_0 (\mu)$.
One can convince oneself, that the DS equations 
eqs.(\ref{eq:6a}) and (\ref{eq:6b}) or (\ref{eq:10a}) and (\ref{eq:10b}) 
are consistent with the renormalization group results 
\cite{LAN,POL,PAG,HIG}:
\begin{equation}
B(p^2)  {\stackrel{p^2 \to \infty}{\sim}}  
  \left [    {m_0(\mu)}  
\left ( \ln {\mu^2 \over \Lambda^2} \right ) ^{d}
\right ]
\left ( \ln {p^2 \over \Lambda^2} \right ) ^{-d}
 +  \left [ 
  {{- 4\pi^2 d}\over 3} 
\langle {\bar q} q \rangle (\mu) 
\left ( \ln {\mu^2 \over \Lambda^2} \right ) ^{-d}
\right ] {1\over {p^2}} 
\left ( \ln {p^2 \over \Lambda^2} \right ) ^{d-1}
\label{eq:15}
\end{equation}
and
\begin{equation}
m_0 ({\tilde{\mu}}) = m_0 (\mu) 
{ 
{ \left ( \ln {{\tilde{\mu^2}} \over \Lambda^2} \right ) ^{d}} \over
{ \left ( \ln {\mu^2 \over \Lambda^2} \right ) ^{d} }
}
\label{eq:16}
\end{equation}
The general behavior of the relationship $m_0(\mu) = m_0(\mu) [B(0)]$ is
the same as without this asymptotic tail.
With the parameter set from refs.\cite{RW,PCR} mentioned above we 
obtain a convergence radius of at least 
$m_{conv} (\mu = 1000 \Lambda) = 1.6 \Lambda$. 
Due to eq.(\ref{eq:16}) this corresponds at a typical
hadronic scale of $\mu \approx 1 \text{GeV}$ 
to a value of $m_{conv} (\mu \approx 1 \text{GeV}) \approx
3 \Lambda$, which lies clearly above the strange quark mass.
\end{enumerate}

Our results can be summarized as follow:
\begin{enumerate}
\item In the framework of the rainbow DS approach, where the form of the 
gluon propagator $D(p^2)$ is taken as input, the convergence radius of 
the chiral perturbation expansion $m_{conv}$ turns out to be crucially
dependent on the infrared behavior of the gluon propagator.
\item $m_{conv}$ gets larger if in the 
IR domain the overall strength of the gluon 
propagator increases or its width decreases.
\item Using a parameterization $D(p^2)$ which provides a good description of
hadronic phenomena at low energies we obtain a value for $m_{conv}$ which
is clearly larger than the strange quark mass and therefore the chiral 
expansion in the strange quark sector converges.
\item Especially we are now able to explain the rather low value of
$m_{conv}$ and therefore the poor convergence of the chiral expansion
in the strange quark sector in case of the NJL or the chiral $\sigma$ model,
where the corresponding gluon propagator is a ``small'' constant in momentum 
space.
Our results indicate that this small value of $m_{conv} \approx 50 \dots
80 \text{MeV}$ in these models arises more likely due to the special form
of a contact interaction than it is a general feature of QCD and the early 
breakdown of the chiral expansion does not occur in a theory with
infrared slavery. 
\end{enumerate}

\acknowledgements
I am very grateful to T.Hatsuda (University of Tsukuba), 
M.Frank (INT, Seattle) and C.Roberts (Argonne National Laboratory) for
numerous useful discussions and comments.
This work has been supported by the Alexander von Humboldt-Stiftung
(Feodeor-Lynen-Programm) and the US Department of Energy 
(grant no. DE-FG06-90ER 40561).

\begin{figure}
\caption{Determination of $m_{conv}$ in the NJL with proper-time
regularization, $\Lambda_{UV} = 634 \text{MeV}$, $G = 30 {\Lambda_{UV}}^2$.}  
\label{fig1}
\end{figure}

\begin{figure}
\caption{Comparison between running coupling constants $\alpha (p)$:
Gaussian form (eq.(\protect\ref{eq:9}) with $\chi = 8 \Lambda$ and $\Lambda = 
200 \text{MeV}$) for various values of $\Delta$ 
and the NJL (eq.(\protect\ref{eq:7b}) with 
$\Lambda_{UV} = 634 \text{MeV}$, $G = 30 {\Lambda_{UV}}^2$).}  
\label{fig2}
\end{figure}

\begin{figure}
\caption{The solution $B(p)$ of the DS equations 
(\protect\ref{eq:10a}) and (\protect\ref{eq:10b})
for a Gaussian running coupling $\alpha (p)$ with various widths $\Delta$
($\chi = 8 \Lambda$, $B(0) = 7.75 \Lambda$, $\Lambda = 200 \text{MeV}$}).
\label{fig3}
\end{figure}

\begin{figure}
\caption{The dependence $m_0 = {m_0}[B(0)]$ obtained from the solutions
of the DS equation 
(\protect\ref{eq:9}) using a Gaussian running coupling compared with the
$\delta$-fct.limes $\Delta = 0$ (eq.(\protect\ref{eq:12a})) and the NJL.
The parameters are the same as in fig.\protect\ref{fig2}.}
\label{fig4}
\end{figure}

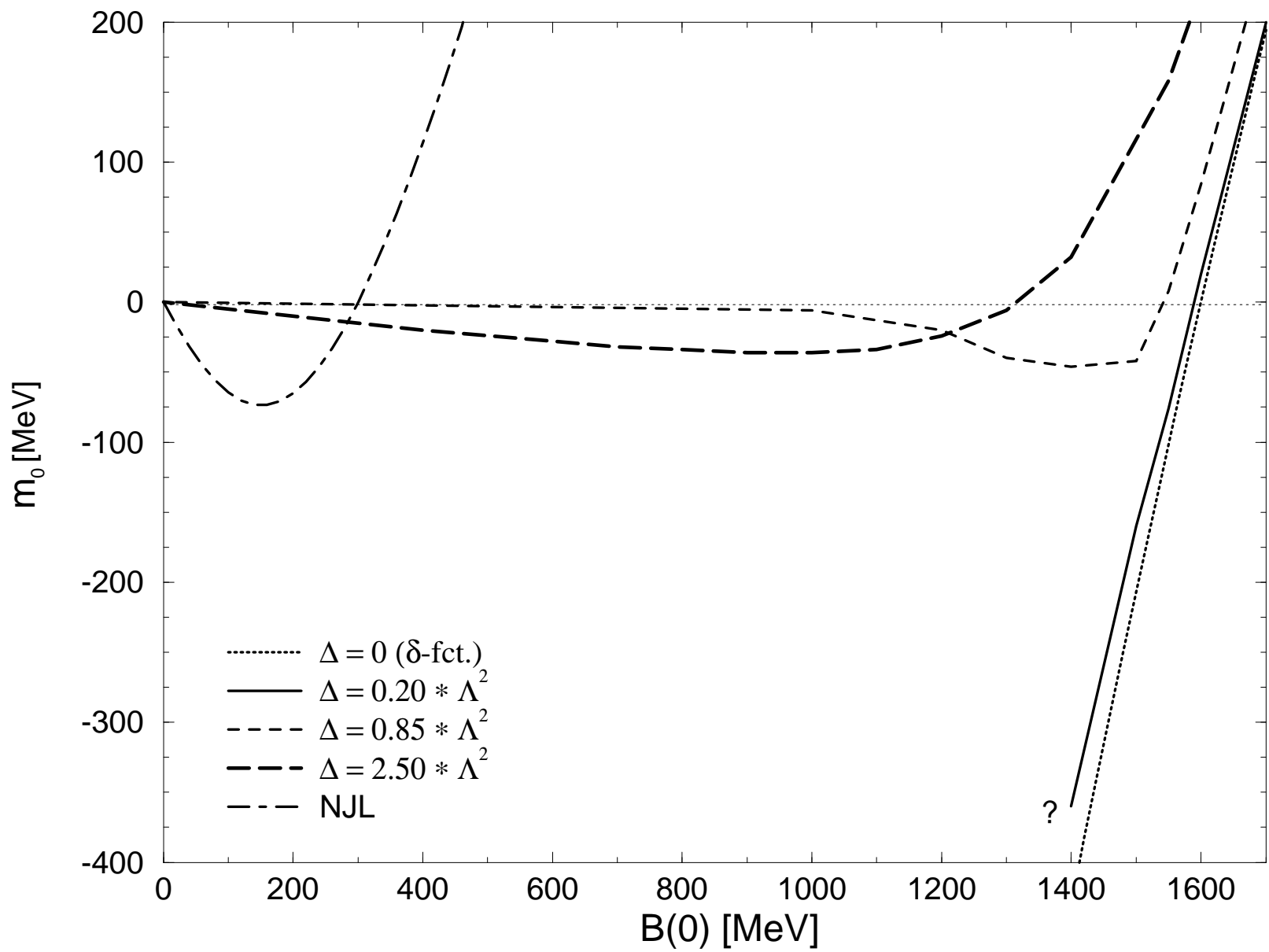
\end{document}